\documentclass[aps,prd,showpacs,amsmath,amssymb,nofootinbib,twocolumn]{revtex4}
\usepackage{bm}
\usepackage{bbm}
\usepackage{color}
\usepackage{graphicx}
\usepackage{times}
\usepackage{psfrag}
\usepackage{mathrsfs}
\newcommand{\Seff}{S_{\text{eff}}}

\DeclareMathOperator{\tr}{tr}

\newcommand{\pol}{\mathfrak{P}}

\DeclareMathOperator{\diag}{diag}

\newcommand{\abs}[1]{\left| #1 \right|}

\newcommand{\bbZ}{\mathbb{Z}}

\newcommand{\vc}[1]{{\bm{#1}}}

\newcommand{\Pol}[1]{\mathfrak{P}_{\vc{#1}}}
\newcommand{\Px}{\mathcal{P}_{\vc{x}}}
\newcommand{\Py}{\mathcal{P}_{\vc{y}}}
\newcommand{\rep}{\mathcal{R}}

\newcommand{\bra}{\langle}
\newcommand{\ket}{\rangle}

\renewcommand{\Re}{\mathrm{Re}\,}
\renewcommand{\Im}{\mathrm{Im}\,}

\newcommand{\sfrac}[2]{{\textstyle \frac{#1}{#2}}}
\newcommand{\dmh}{d\mu_{\mathrm{Haar}}}
\newcommand{\dmr}{d\mu_{\mathrm{red}}}

\newcommand{\pDeriv}[2][]{\frac{\partial #1}{\partial #2}}
\newcommand{\vcx}{\vc{x}}
\newcommand{\vcy}{\vc{y}}
\newcommand{\vcz}{\vc{z}}
\newcommand{\vcv}{\vc{v}}
\newcommand{\vcw}{\vc{w}}
\newcommand{\trP}{\mathcal{P}}
\newcommand{\ev}[1]{\left\langle #1 \right\rangle}
\newcommand{\ord}{\mathcal{O}}

\newcommand{\Nt}{{N_\mathrm{t}}}
\newcommand{\nN}[1]{\left\langle #1 \right\rangle}
\newcommand{\onN}[1]{\left[ #1 \right]}
\newcommand{\qnN}[1]{\left( #1 \right)}
\newcommand{\coco}{\mathrm{c.c.}}

\begin{document}

\title{Inverse Monte-Carlo determination of effective lattice
models for SU(3) Yang-Mills theory at finite temperature}

\author{Christian Wozar, Tobias Kaestner and Andreas Wipf}
\affiliation{Theoretisch-Physikalisches Institut,
Friedrich-Schiller-Universit{\"a}t Jena, Max-Wien-Platz 1, 07743
Jena, Germany}

\author{Thomas Heinzl}
\affiliation{School of Mathematics and Statistics, University of
Plymouth, Drake Circus, Plymouth, PL4 8AA, United Kingdom}

\begin{abstract}

This paper concludes our efforts in describing $SU(3)$-Yang-Mills
theories at different couplings/temperatures in terms of effective
Polyakov-loop models. The associated effective couplings are
determined through an inverse Monte Carlo procedure based on novel
Schwinger-Dyson equations that employ the symmetries of the Haar
measure. Due to the first-order nature of the phase transition we
encounter a fine-tuning problem in reproducing the correct
behavior of the Polyakov-loop from the effective models. The
problem remains under control as long as the number of effective
couplings is sufficiently small.

\end{abstract}

%11.15.Ha Lattice gauge theory
%11.15.Me Strong-coupling expansions
%11.10.Wx Finite-temperature field theory
%12.40.Ee Statistical models
\maketitle

\section{Introduction}\label{sec:intro}

Since the pioneering papers of Polyakov \cite{Polyakov:1978vu} and
Susskind \cite{Susskind:1979up} the confinement-deconfinement
phase transition in finite temperature Yang-Mills theory has
become a thoroughly studied phenomenon. This is particularly true
for the `standard' groups $SU(2)$ and $SU(3)$, but more `exotic'
groups like $G(2)$ have recently come into focus as well, see
e.g.\ \cite{Holland:2003jy,Greensite:2006sm}. Due to the
nonperturbative nature of the problem progress has mainly been
achieved via brute force lattice computations.

Nevertheless, it would be helpful to have a simpler and more
intuitive understanding of the physics involved. The principal
tool for this purpose is the construction and subsequent analysis
of effective models. Such attempts also have quite some history
going back to the works of Svetitsky and Yaffe
\cite{Yaffe:1982qf,Svetitsky:1982gs} as well as Polonyi and
Szlachanyi \cite{Polonyi:1982wz}. The basic idea (in the spirit of
Landau and Ginzburg) is to use the order parameter of the
transition, the Polyakov loop, as a collective degree of freedom
and formulate effective theories in terms of it. For gauge groups
$SU(N)$ the rationale behind that is the Svetitsky-Yaffe
conjecture \cite{Yaffe:1982qf,Svetitsky:1982gs} which states that
the Yang-Mills finite-temperature transition in dimension $d+1$ is
described by an effective spin model in $d$ dimensions with
short-range interactions\footnote{The reasoning involved strongly
relies on center symmetry. The study of more exotic Lie groups
(which may not even have a nontrivial center) suggests that the
size of the gauge group is also important \cite{Holland:2003kg}.}.
These ideas have initially been taken up in terms of strong
coupling expansions
\cite{Polonyi:1982wz,Green:1983sd,Ogilvie:1983ss} yielding Ising
type spin models with an effective coupling $\lambda(\beta)$ where
$\beta$ denotes the Yang-Mills-Wilson coupling. A review of early
work in this context may be found in \cite{Svetitsky:1985ye}. For
an overview of more recent developments we refer the reader to
\cite{Holland:2000uj}.

Early on, it has also been attempted to obtain these effective
models, that is their couplings (being the `weights' of the
included operators) nonperturbatively via lattice methods. In
\cite{Creutz:1984fj,Gocksch:1984ih} Creutz's microcanonical demon
method \cite{Creutz:1983ra} has been employed for $SU(2)$. An
alternative method based on Schwinger-Dyson equations (SDEs) and
dubbed ``inverse Monte Carlo'' (IMC) was developed soon after
\cite{Falcioni:1985bh,Gonzalez-Arroyo:1986ck} and applied to both
$SU(2)$ \cite{Deckert:1987we,Gonzalez-Arroyo:1987pz} and $SU(3)$
\cite{Fukugita:1989yb,Moore:1989qw}. Since then the IMC approach
to lattice gauge theories has largely been dormant with only a few
exceptions \cite{Hasenbusch:1994ne,Svetitsky:1997du}.

Inspired by the success of Polyakov loop models
\cite{Pisarski:2000eq,Pisarski:2001pe,Dumitru:2003hp} we have
recently reinvestigated the feasibility of IMC for the
confinement-deconfinement phase transition in a series of papers.
Our numerical approach has consistently been complemented by
analytical attempts such as strong coupling expansions and
mean-field approximations. For the second-order $SU(2)$ transition
our results may be found in \cite{Dittmann:2003qt} and
\cite{Heinzl:2005xv}. By including up to 14 operators and 3
different group representations we were able to reproduce suitable
Yang-Mills observables to a reasonable accuracy. The same is true
for the analytically known asymptotic behavior of the effective
couplings as a function of $\beta$. In \cite{Wozar:2006fi} we have
started to investigate effective models for $SU(3)$ which are
generalisations of the 3-states Potts model. The critical behavior
of these is an interesting subject in its own right. We have found
a very rich phase structure with first and second order
transitions between symmetric, ferromagnetic and
anti-ferromagnetic phases. In addition there seems to be a
tricritical point rendering mean-field theory approximately exact
in its vicinity \cite{Wozar:2006fi,Wipf:2006wj}

In relating the effective models to $SU(3)$ Yang-Mills via IMC one
expects to encounter new difficulties. The first technical problem
to overcome is to find the SDEs which are less straightforward
than for $SU(2)$ as the $SU(3)$ group manifold no longer is a
sphere. This problem has been solved in \cite{Wozar:2006dp} and
\cite{Uhlmann:2006ze}. As the $SU(3)$ phase transition is of
(weak) first order, hence not continuous, the determination of the
effective couplings might require fine-tuning raising the question
of stability of the solutions. This will be one of the main issues
to be addressed in what follows.

The remainder of the paper is organized as follows. In
Section~\ref{sec:sota} we suggest different effective actions as
candidates for describing the Polyakov loop dynamics of Yang-Mills
theory. In Section~\ref{sec:sdIMC} we explain how to obtain the
effective couplings via two alternative sets of SDEs and
subsequent IMC. Our numerical results are presented in
Section~\ref{sec:numRes}. Section \ref{sec:SO} concludes our
discussion with a summary and outlook. Some technicalities are
deferred to Appendices \ref{app:SLLIST}--\ref{app:LAMBDA-KAPPA}.

% ========================================
\section{Effective actions}
\label{sec:sota}
% ========================================

We begin by recalling the lattice definition of the untraced
Polyakov loop in the group representation $\rep$,
\begin{equation} \label{POL}
\rep(\Pol{x}) \equiv \prod_{t=1}^{\Nt} \rep(U_{t\vc{x}}) \; ,
\end{equation}
where $\rep(U_{t\vc{x}})$ is the temporal link at time slice $t$
and position $\vc{x}$ in representation $\rep$. Any irreducible
representation of $SU(3)$ is labeled by two integers, $\rep =
\rep_{pq}$, which, in flavor language, count the number of quarks
and antiquarks needed to construct the multiplet associated with
$\rep_{pq}$. The basic building blocks of our effective actions
are the group characters $\chi_\rep$ associated with the
representation $\rep$, that is the traces of the Polyakov loop
\eqref{POL},
\begin{equation}
  \chi_\rep (\trP) \equiv \chi_{pq} (\trP) \equiv \tr \rep_{pq} (\pol) \; .
\end{equation}
Note that these only depend on the traced Polyakov loop in the fundamental
representation, $\trP = \tr \pol$. The trivial character is of course
$\chi_{00} = 1$ while the first nontrivial ones correspond to the
(anti-)fundamental representations and yield just the standard traced Polyakov
loop (and its complex conjugate),
\begin{equation}
  \chi_{10} (\trP) = \trP \; , \quad  \chi_{01} (\trP) = \trP^* \; .
\end{equation}
Under a center transformation, the characters transform as %
\begin{equation}
  \chi_{pq} \to z^{p-q} \chi_{pq}  \; , \quad z \in \mathbb{Z}_3 \; .
\end{equation}
Center symmetry is then sufficient to determine the operator content of the
effective action if we restrict to nearest-neighbor (NN) interactions. In terms
of group characters one finds the general form
\begin{equation}\label{DUMITRU}
  S_{\mathrm{eff}}[\chi] = \sum_{\substack{ \bra \vc{x} \vc{y}\ket , \, pq,
  \, p'q' \\ p + p' = q + q' \bmod  3}} \lambda_{pq, p'q'} \,
  \chi_{pq} (\Px) \, \chi_{p'q'} (\Py)  \; ,
\end{equation}
where the sum over representations is constrained by center
symmetry. Expressing the characters explicitly in terms of the
Polyakov loop $\trP$ one easily recognizes  \eqref{DUMITRU} as the
action suggested by Dumitru et al. \cite{Dumitru:2003hp}. Their
`potential terms', built from single center symmetric characters
located at single sites appear whenever the second adjacent
character is trivial, $\chi_{00} = 1$. In this case the typical
hopping terms connecting NN sites `degenerate' into ultra-local
terms, the one of lowest dimension being the `octet loop'
contribution, $\lambda_{11,00} \, \chi_{11}$ as indeed $1 + 0 = 1
+ 0 \bmod 3$. One expects that the couplings $\lambda_{pq, p'q'}$
decrease with increasing representation labels, $p$, $q$, $p'$ and
$q'$, hence that representations of low dimension,
\begin{equation}
d_{pq} = \sfrac{1}{2} (p+1)(q+1)(p+q+2) \; ,
\end{equation}
dominate the effective action. To simplify our notation we will
henceforth write the action \eqref{DUMITRU} (and generalizations
thereof) as a series of the form
\begin{equation} \label{S-LAMBDA}
\Seff = \sum_i \lambda_i \, S_i \; ,
\end{equation}
where up to 16 different terms $S_i$ will be taken into consideration, albeit
not necessarily within one and the same ansatz. A list of the operators $S_i$
may be found in Appendix \ref{app:SLLIST} where we allow for next-to-NN (NNN)
couplings in addition. It is easy to check that each of the terms given
satisfies the selection rules for the representation labels necessary for
center symmetry.

It turns out (in hindsight) that the \textit{ans\"atze}
\eqref{DUMITRU} or \eqref{S-LAMBDA} contain more freedom of choice
than actually required which makes the inverse Monte Carlo
routines less efficient (see below). To further constrain this
freedom we generalise to $SU(3)$ the strong-coupling approach
introduced by Billo et al. \cite{Billo:1996wv} which we already
have successfully applied to $SU(2)$ \cite{Heinzl:2005xv}. The
basic building blocks are then given by the center symmetric
operators connecting NN sites,
\begin{equation}
  S_{\rep, \ell} \equiv \chi_\rep (\trP_\vcx) \chi_\rep^* (\trP_{\vcy}) + \coco
  \; , \quad \ell \equiv \nN{\vcx\vcy} . \label{SLINK}
\end{equation}
The strong coupling expansion then replaces the ansatz \eqref{DUMITRU} by the
following somewhat more complicated expression \cite{Wozar:2006dp},
\begin{equation} \label{SSC}
  S_{\mathrm{eff}} = \sum_r \sum_{\rep_1 \ldots  \rep_r} \sum_{\ell_1 \ldots \ell_r}
  c_{\rep_1 \ldots \rep_r}^{\ell_1 \ldots  \ell_r}(\beta)  \prod_{i=1}^r
  S_{\rep_i, \ell_i} \; ,
\end{equation}
where $r$ counts the number of link operators \eqref{SLINK}
contributing at each order. The coefficients $c_{\rep_1 \ldots
\rep_r}^{\ell_1 \ldots \ell_r}$ are the couplings between the
operators $S_{\rep_i,\ell_i}$ from \eqref{SLINK} sitting at NN
links $\ell_i \equiv \nN{\vcx_i,\vcy_i}$ in representation
$\rep_i$. The effective action defined in \eqref{SSC} hence
describes a network of link operators of the type \eqref{SLINK}
that are collected into (possibly disconnected) `polymers'
contributing with `weight' $c_{\rep_1 \ldots \rep_r}^{\ell_1
\ldots \ell_r}$. Again, one expects the `weights' or couplings to
decrease as the dimensions of the representations and the
inter-link distances involved increase. In a strong coupling
(small-$\beta$) expansion truncated at $\ord(\beta^{k \Nt})$ one
has $r \le k$ and the additional restriction $|\rep_1| + \cdots +
|\rep_r| < k$ with $|\rep| \equiv p+q $ for a given representation
$\rep$.

To lowest order $\beta^\Nt$ one finds the universal effective action
\cite{Polonyi:1982wz}
\begin{multline}
\Seff = c_{10} \sum_{\nN{\vcx\vcy}} S_{10, \nN{\vcx\vcy}}
\equiv \kappa_1 \sum_{\nN{\vcx\vcy}} (
\chi_{10}(\trP_\vcx)\chi_{01}(\trP_\vcy) + \coco ) \\
\equiv \kappa_1
\sum_{\nN{\vcx\vcy}} (\trP_\vcx  \trP_\vcy^*  +  \trP_\vcx^* \trP_\vcy ) \; ,
\end{multline}
which is just a single hopping term connecting Polyakov loops at
NN sites. This is reminiscent of a generalised Ising model or,
more appropriately, a three-state Potts model \cite{Potts:1952}.
As mentioned before the study of these models is interesting in
its own right \cite{Wozar:2006fi} but will not be the topic of the
present paper which focuses on the relation between the effective
actions and Yang-Mills theory.

Again, the most general representation \eqref{SSC} is not too
illuminating and we will therefore adopt the notation
\begin{equation} \label{S-KAPPA}
  \Seff = \sum_a \kappa_a I_a \; .
\end{equation}
A list of the leading NN and NNN action terms $I_a$ can be found in
Appendix~\ref{app:SKLIST}.

In principle, without any truncations, the effective actions
\eqref{S-LAMBDA} and \eqref{S-KAPPA} have to coincide although the
operator bases used are different. Accordingly, there is a linear
relationship between the couplings $\lambda_i$ and $\kappa_a$.
However, as the ordering principles for the two \textit{ans\"atze}
are not the same the relation between the couplings \textit{after}
truncation is \textit{not} one-to-one but rather of the form
\begin{equation} \label{LAMBDA-KAPPA}
\lambda_i = K_{ia} \kappa_a \; ,
\end{equation}
Typically, for a given truncation of the $\lambda$-action \eqref{S-LAMBDA} the
range of the index $i$ is larger than that of $a$, i.e.\ there are more
$\lambda$'s than $\kappa$'s. Accordingly, the matrices $(K_{ia})$ are rectangular with
$a < i$ and integer entries. For this reason it is calculationally often more
efficient to work with the $\kappa$-action \eqref{S-KAPPA} and reobtain the
$\lambda_i$ via \eqref{LAMBDA-KAPPA}. Our standard choices for the numerical
matrices $(K_{ia})$ corresponding to different truncations of the effective
actions may be found in Appendix~\ref{app:LAMBDA-KAPPA}. As the operators $S_i$
appearing in \eqref{S-LAMBDA} are more intuitive and resemble generalised spin
terms we have decided, for the sake of brevity, to present results only for the
$\lambda$'s in this paper.

% ========================================
\section{Schwinger-Dyson equations and inverse Monte-Carlo}
\label{sec:sdIMC}
% ========================================

In this section we shortly recapitulate the $SU(3)$
Schwinger-Dyson equations (SDEs) that have recently been derived
in \cite{Wozar:2006dp} and \cite{Uhlmann:2006ze}. They will be the
main tool to relate the effective actions \eqref{S-LAMBDA} and
\eqref{S-KAPPA} to Yang-Mills theory. Our numerical approach
benefits from the fact that there are two independent versions of
SDEs which in the end, however, should yield equivalent results.
The first type of equations is based on an integral identity which
is more algebraic in nature than the second type which follows
from geometrical considerations.

% ========================================
\subsection{Algebraic SDEs}
% ========================================

It is useful to parameterize the diagonalized, untraced Polyakov
loop by means of two angular variables, $\phi_1$ and $\phi_2$,
\begin{equation} \label{ANGLEREP}
   \pol (\phi_1,\phi_2) = \begin{pmatrix}
    e^{i\phi_1} & 0 & 0 \\
    0 & e^{i\phi_2} & 0 \\
    0 & 0 & e^{-i(\phi_1+\phi_2)}
    \end{pmatrix} \; ,
\end{equation}
with values in a fundamental region given by the restrictions
\begin{equation} \label{FUNDDOM}
  \phi_1 \le \phi_2 \le (-\phi_1-\phi_2) \bmod 2\pi\;,\quad 0\le \phi_i<2\pi \; .
\end{equation}
As a result, the reduced Haar measure acquires the form
\begin{equation} \label{eq:DMUPHI}
  d\mu_\mathrm{red} = J^2 d\phi_1 d\phi_2
\end{equation}
with a nontrivial Jacobian that may either be expressed in terms of characters,
\begin{equation}
  J^2 = 15 - 6\chi_{11} + 3\chi_{30} + 3\chi_{03} - \chi_{22} \; ,
\end{equation}
or in terms of the trace $\trP$ of \eqref{ANGLEREP},
\begin{equation}
  J^2 = 27 - 18 \trP \trP^* + 4\trP^3 + 4{\trP^*}^3 - \trP^2 {\trP^*}^2 \; .
\end{equation}
Using the latter variable leads to the remarkable algebraic identity
\begin{equation} \label{3J}
   d\mu_\mathrm{red} (\phi_1, \phi_2) = J^2 \, d\phi_1 \, d\phi_2 = J \, d\trP \,
   d\trP^* = d\mu_{\mathrm{red}} (\trP , \trP^*) \; ,
\end{equation}
such that one can trade $J^2$ for its square root $J$. It is a fact of life
that any function $f = f(\trP, \trP^*)$ vanishing on the boundary $\partial
\Omega$ of a region $\Omega$ satisfies the integral identity
\begin{equation} \label{INT1}
  \int_\Omega d\trP \, d{\trP^*} \, \partial_\trP f = 0 \; ,
\end{equation}
which is reminiscent of integration by parts on the real line, $\int dx \, f'(x)
= 0$ (for functions $f$ vanishing at infinity). For our purposes we
make the particular choice
\begin{equation}
  f(\trP,{\trP^*}) = J^3 g(\trP,{\trP^*})
\end{equation}
with arbitrary $g$ and integrate over the domain of $\trP$ given implicitly via
\eqref{FUNDDOM}. This is consistent with the general identity \eqref{INT1} as the reduced Haar
measure vanishes at the boundary $\partial \Omega$ of the fundamental region
\eqref{FUNDDOM}.  Hence, \eqref{INT1} specializes to
\begin{equation}\label{INT2}
  0 = \int d\mu_\mathrm{red}(\trP_\vcz , \trP^*_\vcz) \left(
  \frac{3}{2} \pDeriv[J_\vcz^2]{\trP_\vcz} \, g + J_\vcz^2 \,
  \pDeriv[g]{\trP_\vcz} \right) \; ,
\end{equation}
where we have used the reduced Haar measure from \eqref{3J} and reinstated the
dependence on the lattice site chosen to be $\vcz$. The derivative of $J^2$ can
actually be worked out with the result %
\begin{equation}
  \pDeriv[J^2]{\trP} = 12\trP^2 -2\trP{\trP^*}^2 -18{\trP^*} \; .
\end{equation}
To actually obtain genuine SDEs we have to introduce the usual
Boltzmann factor $\exp(-\Seff)$. This is done by choosing a special
function $g$, namely
\begin{equation} \label{GX}
  g_\vcx \equiv \pDeriv[h]{\trP^*_\vcx} \exp(-\Seff) \; ,
\end{equation}
with another function $h(\trP,{\trP^*})$ required to be $\bbZ_3$ invariant,
\begin{equation} \label{Z3INV}
   h (\trP, \trP^*) = h (z \trP, z^* \trP^*) \; , \quad z \in \bbZ_3 \; .
\end{equation}
Plugging in the ansatz \eqref{S-LAMBDA} for the $\lambda$-action the
$\trP$-derivative of \eqref{GX} needed for \eqref{INT2} becomes
\begin{equation}
  \pDeriv[g_\vcx]{\trP_\vcz} = \left( h_{,\trP^*_\vcx,\trP_\vcz}-
  \sum_i \lambda_i h_{,\trP^*_\vcx} S_{i, \trP_\vcz}\right) e^{-\Seff} \;
  ,
\end{equation}
where the commas denote differentiation with respect to the subsequent argument.
Functional integration of \eqref{INT2} with the measure $\mathscr{D}\mu_{\mathrm{red}}
\equiv \prod_{\vcz} d\mu_\mathrm{red}(\trP_\vcz , \trP^*_\vcz)$ finally yields
the desired SDEs,
\begin{equation}\label{ASDE}
  0 = \ev{\frac{3}{2}\pDeriv[J_\vcz^2]{\trP_\vcz} h_{,\trP^*_\vcx}
  + J_\vcz^2 h_{,\trP^*_\vcx,\trP_\vcz}}-\sum_i \lambda_i
  \ev{J_\vcz^2 h_{,\trP^*_\vcx} S_{i, \trP_\vcz} } \; ,
\end{equation}
which comprise a linear system for the effective couplings $\lambda_i$,
generalising analogous results for $SU(2)$ \cite{Dittmann:2003qt,Heinzl:2005xv}.
The experience gained there prompts us to choose the function $h$ from the
operators $S_i$ in the ansatz for the $\lambda$-action \eqref{S-LAMBDA}. Any
index $i$ then yields an independent equation. In addition, this choice
automatically satisfies the criterion \eqref{Z3INV} of $\bbZ_3$ invariance.

On top of that we will vary the sites $\vcx$ and $\vcz$, in particular the
distance $d \equiv \abs{\vcx-\vcz}$ between them. On a lattice with spatial
extent $N_\mathrm{s}$ this implies a range of distances $d \in \{0, \ldots,
\left\lfloor N_\mathrm{s}/2 \right\rfloor\}$ where $\lfloor x \rfloor$ denotes
the largest integer $\le x$. For $N$ different operators $S_i$ we thus obtain
$N$ independent equations for each distance $d$.

% ========================================
\subsection{Geometrical SDEs}
% ========================================

For any function $f = f(U)$ on a Lie group we define its left derivative in the
direction of the generator $T^a$ via
\begin{equation}
  L_a f(U) \equiv \frac{d}{dt} f \left. \left( e^{t T^a} U \right)
  \right|_{t=0} \; .
\end{equation}
Left invariance of the Haar measure implies a symmetry relation somewhat
analogous to \eqref{INT1},
\begin{equation} \label{eq:DMH}
  \int \dmh (U) \, L_a f(U) = 0 \; ,
\end{equation}
which will serve as a master identity generating all SDEs. As in
the previous subsection we would like to integrate over the
reduced Haar measure $d\mu_\mathrm{red}$ only. Thus, we want the
integrand $L_a f(U)$ to be a \emph{class} function. If $G$ is such
a class function it only depends on the fundamental group
characters, $\chi_F (U) \equiv \tr (\rep_F(U))$, where $\rep_F(U)$
denotes a fundamental representation of the group element $U$,
i.e.\ $\mathbf{3}$ or $\bar{\mathbf{3}}$ for $SU(3)$. For the
particular choice $f = g L_a \chi_p$, with $g$ an arbitrary class
function and $\chi_p$ a fundamental character, the integrand in
\eqref{eq:DMH} indeed becomes a class function so that we end up
with the `reduced' integral
\begin{equation}
  \int \dmr \, L_a (g L^a \chi_p) = 0 \; .
\end{equation}
For $SU(3)$ we obviously choose the fundamental character $\chi_p$ as the trace
of the Polyakov loop in the fundamental representation, $\chi_p \equiv
\chi_{10} \equiv  \trP$ and $g \equiv h \, e^{-\Seff}$, slightly different from
\eqref{GX}. The left derivatives are worked out as follows \cite{Wozar:2006dp,Uhlmann:2006ze},
\begin{equation}
  \begin{split}
  L_a(L^a(\trP)) &= -\frac{16}{3} \trP\;,\\
  L_a(\trP)L^a(\trP) &= 4\trP^*-\frac{4}{3}\trP^2 \;,\\
  L_a(\trP)L^a(\trP^*) &= 6-\frac{2}{3}\abs{\trP}^2 \; ,
  \end{split}
\end{equation}\\[1em]
and only depend on the traced Polyakov loop as they should. Finally, to obtain
feasible equations we choose $h$ among the $\trP$-derivatives
of the $S_i$,  $h = {S_i}_{,\trP_\vcx}$ implying the following set of
geometrical SDEs,
\begin{widetext}
\begin{multline} \label{GSDE}
  0 = \left\langle-\frac{16}{3} \trP_\vcz S_{i,\trP_\vcx} + (4\trP^*_\vcz-\frac{4}{3}\trP^2_\vcz)
  S_{i,\trP_\vcx,\trP_\vcz} +
  (6-\frac{2}{3}\abs{\trP_\vcz}^2) S_{i,\trP_\vcx,\trP^*_\vcz}\right\rangle  \\
  - \sum_j \lambda_j \left\langle (4\trP^*_\vcz-\frac{4}{3}\trP^2_\vcz)
  S_{i,\trP_\vcx} S_{j,\trP_\vcz}
 + (6-\frac{2}{3}\abs{\trP_\vcz}^2) S_{i,\trP_\vcx}
  S_{j,\trP^*_\vcz} \right\rangle \;,
\end{multline}
\end{widetext}
where, again, the dependence on the lattice site $\vcz$ has been made explicit.

% ========================================
\subsection{Normalisation}
% ========================================

We have seen that, for every pair of lattice sites $\vc{x}$ and
$\vc{y}$, we end up with a linear system of equations for the
couplings of the effective theory. Since on the lattice we have
both translational and (discrete) rotational symmetry it is
sufficient to consider different distances $d =
\abs{\vc{x}-\vc{y}}$ only.  These serve as a label for our sets of
equations which in a condensed matrix notation may be written as
\begin{equation} \label{EQLAMBDA}
  M_d \, \vc{\lambda} = \vc{b}_d \; .
\end{equation}
If we assume a total of $N$ unknown couplings collected into the
vector $\vc{\lambda}$ we have, by construction of the SDEs, an $N
\times N$ matrix $M_d$ and an inhomogeneity $\vc{b}_d$, hence an
independent system of equations, for each distance $d$. The
off-diagonal entries of $M_d$ and the vector $\vc{b}_d$ are
typically complex but the couplings $\vc{\lambda}$ have to be
real. We therefore distinguish between real and imaginary parts of
\eqref{EQLAMBDA} for different $d$ and group them together into
the equations
\begin{equation} \label{OVEREQLAMBDA}
  \begin{pmatrix}
  \Re M_0 \\ \Im M_0  \\ \Re M_1 \\ \Im M_1  \\ \vdots \\ \Im M_{\lfloor
  N_s/2\rfloor} \end{pmatrix} \vc{\lambda} = \begin{pmatrix}
  \Re \vc{b}_0 \\ \Im \vc{b}_0 \\ \Re \vc{b}_1 \\ \Im \vc{b}_1  \\ \vdots \\ \Im \vc{b}_{\lfloor
  N_s/2 \rfloor} \end{pmatrix} \; .
\end{equation}
This constitutes an overdetermined linear system of $2N \times (\lfloor N_s/2
\rfloor + 1)$ equations for the $N$ unknown couplings $\vc{\lambda}$. In
principle, this can be solved by standard least-square methods.

However, this procedure is hampered by a few technical pitfalls. Since the order
parameter for the confinement-deconfinement transition is driven by the
long-range behavior of the lattice system we have to take into account the fact that
equations associated with different distances $d$ enter \eqref{OVEREQLAMBDA}
with different multiplicities. On a three-dimensional lattice this entails that an
equation for distance $d>0$ has multiplicity
\begin{equation} \label{MULT}
  m_d \equiv (2d+1)^3-(2d-1)^3
\end{equation}
while equations for $d=0$ only appear once. To account for this
mismatch we reweight the coefficients of $\vc{b}_d$ and $M_d$ (for
$d>0$) with a factor $\sqrt{m_d}$.

Another problem are the large condition numbers of the matrices. To cope with
this we employ a simple form of normalization based on the diagonal elements of
the matrix $M_0$. For the algebraic SDEs the diagonal entries $M_{0,ii}$ of
$M_0$ dominate the linear system. For this reason we construct a new diagonal
matrix $N_0$ from the $M_{0,ii}$ according to
\begin{equation}
  N_0  \equiv \diag \bigl( \Re(M_{0,11})^{-1/2}, \ldots ,
  \Re(M_{0,NN})^{-1/2} \, \bigr)  \; .
\end{equation}
By means of a similarity transformation with $N_0$ the real parts of the
$M_{0,ii}$ may be transformed to unity. As a result \eqref{OVEREQLAMBDA} becomes
\begin{equation} \label{CONDEQLAMBDA}
  (N_0 M_d N_0) (N_0^{-1} \vc{\lambda}) \equiv (N_0 M_d N_0) \, \vc{\mu} = N_0
  \, \vc{b}_d \; .
\end{equation}
Typically, this new system of equations is better conditioned
which increases the stability of the results.

In our numerical calculations we have both used the improvement \eqref{CONDEQLAMBDA} in
condition numbers and the reweighting factors given by the square root of
\eqref{MULT}.

% ========================================
\section{Numerical results}
\label{sec:numRes}
% ========================================

The IMC method basically amounts to solving the SDEs \eqref{ASDE}
or \eqref{GSDE}, the crux being the evaluation of the expectation
values $\langle \ldots \rangle$ in the \emph{microscopic}
ensemble. In our case this is given by a sufficient number of
$SU(3)$ Yang-Mills configurations generated by standard MC
routines \cite{Cabibbo:1982zn,Creutz:1980zw,Kennedy:1985nu}. In what
follows we will consider three
\textit{ans\"atze}, one with five NN couplings and two more
general ones which either contain NN terms in higher
representations or additional NNN couplings. In a strong coupling
expansion (which is rationale for the $\kappa$-actions) both the
five leading NN terms as well as the NNN ones would be of order
$\beta^{2\Nt}$ while the extended NN ones would be
$\ord(\beta^{3\Nt})$.

\subsection{NN couplings}

Before one determines the effective couplings corresponding to
Yang-Mills it is prudent to check if the SDEs \eqref{ASDE} and
\eqref{GSDE} derived above are consistent within the effective
theories themselves. To test for that we have first simulated an
effective theory with five fixed input couplings and tried to
reproduce them via IMC. In Table \ref{tab:consistency1} we have
listed the outcome of this testing procedure. One notes that the
couplings $\lambda_{i,\mathrm{IMC}}$ determined via IMC coincide
with the chosen input couplings $\lambda_{i,\mathrm{input}}$ to an
accuracy of about 2\%, both for the algebraic and geometrical SDEs.
This tells us two things, first that our SDEs \eqref{ASDE} and
\eqref{GSDE} are both correct and, second, that IMC works extremely
well for the effective theories.

\begin{table}[h]
\caption{\label{tab:consistency1}IMC consistency check for the NN
$\lambda$-action including five couplings $\lambda_1, \ldots ,
\lambda_5$.}
\begin{ruledtabular}
\begin{tabular}{lcccc}
 & \multicolumn{2}{c}{algebraic} & \multicolumn{2}{c}{geometrical} \\
  $i$ & $\lambda_{i,\mathrm{input}}$ & $\lambda_{i,\mathrm{IMC}}$ &
  $\lambda_{i,\mathrm{input}}$ & $\lambda_{i,\mathrm{IMC}}$ \\ \hline
  $1$ & $-0.0100$ & $-0.0101(1)$ & $-0.0100$ & $-0.0101(1)$ \\
  $2$ & $0.0060$ & $0.0059(1)$ & $0.0060$ & $0.0060(1)$ \\
  $3$ & $-0.0050$ & $-0.0051(1)$ & $-0.0050$ & $-0.0051(1)$ \\
  $4$ & $0.0080$ & $0.0079(1)$ & $0.0080$ & $0.0080(1)$ \\
  $5$ & $-0.0060$ & $-0.0059(4)$ & $-0.0060$ & $-0.0060(2)$ \\
\end{tabular}
\end{ruledtabular}
\end{table}
Having thus gained confidence in the validity of our IMC approach
and its implementation we can move on to apply it to our objective
namely to determine effective actions reproducing Yang-Mills
thermodynamics, in particular the deconfinement phase transition.
The first question we want to consider is whether the effective
couplings $\lambda_i$ viewed as functions of the Wilson coupling
$\beta$ are sensitive to the phase transition.
\begin{figure}[h]
\includegraphics{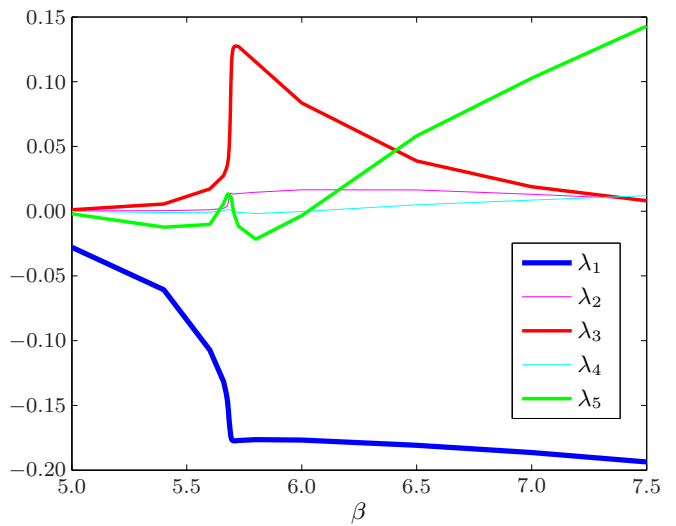}
\caption{\label{fig:couplings4} Behavior of the couplings
$\lambda_1, \ldots, \lambda_5$ (appearing in the NN
$\lambda$-action) as a function of the Wilson coupling $\beta$.
(IMC based on algebraic SDEs.)}
\end{figure}
The answer turns out to be affirmative: Fig.~\ref{fig:couplings4}
clearly shows a rather drastic change in the behavior of the
couplings at a value of $\beta_\lambda \simeq 5.69$ for all
$\lambda_i$, $i = 1, \ldots , 5$. Across the transition, i.e.\ in
both phases the dominant coupling is the `fundamental' one,
$\lambda_1$, followed by the octet couplings $\lambda_3$ and
$\lambda_5$. The latter is actually a `potential' coupling in the
sense of \cite{Dumitru:2003hp} as it multiplies the center
symmetric single-site octet character $\chi_{11}$ (see
App.~\ref{app:SLLIST}). The couplings $\lambda_2$ and $\lambda_4$
are clearly subdominant.

\begin{figure}[h]
\includegraphics{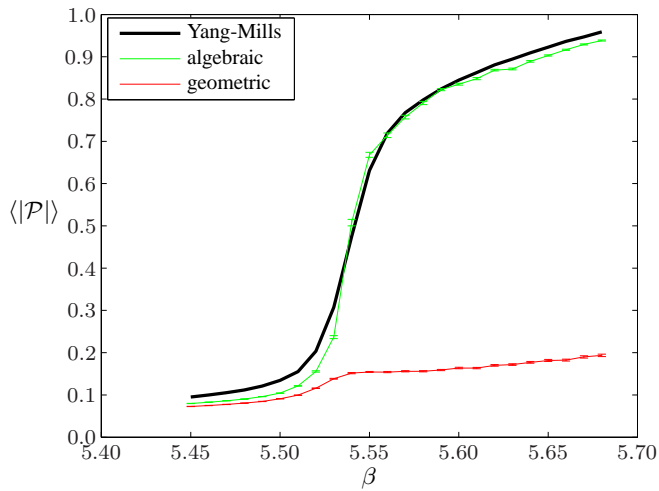}
\caption{\label{fig:eqTypeComparison} Comparison between Yang-Mills and effective
Polyakov loops for a lattice of size $8^3\!\times\! 3$ and five NN couplings
$\lambda_1, \ldots, \lambda_5$.}
\end{figure}

The natural observable to address is, of course, the Polyakov loop
which serves as the order parameter of the first-order $SU(3)$ phase
transition. In Fig.~\ref{fig:eqTypeComparison} we compare the
effective and Yang-Mills Polyakov loops for a relatively small
lattice of size $8^3\!\times\! 3$ where the would-be discontinuities
(in infinite volume) of the transition are still fairly smooth.
Somewhat surprisingly it is only the algebraic SDEs which reproduce
the behaviour of the Yang-Mills Polyakov loop reasonably well. The
geometrical SDEs, on the other hand, fail to do so, at least in the
region just above the transition point. This is a first hint that
there is some inherent instability in the IMC procedure -- in
particular if the geometrical SDEs are used.

\begin{figure}[h]
\includegraphics{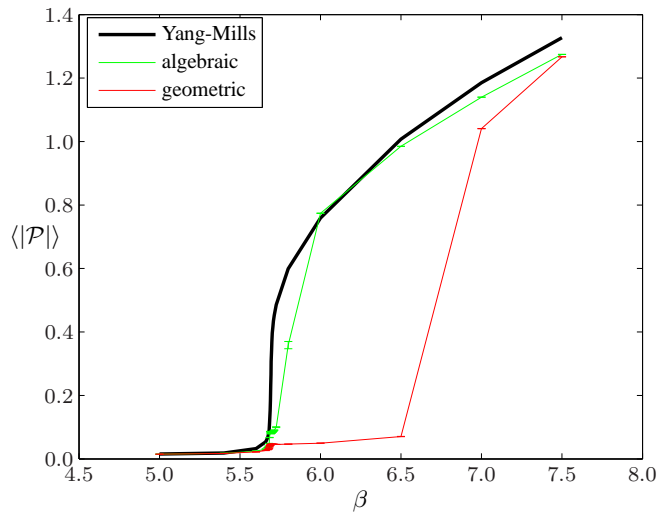}
\caption{\label{fig:polFit16} Comparison between Yang-Mills and effective
Polyakov loops for a lattice of size $16^3 \times
4$ and five NN couplings $\lambda_1, \ldots \lambda_5$. }
\end{figure}

If we move on to larger lattices where the jump of the order
parameter at the critical coupling becomes more pronounced one
finds the behavior displayed in Fig.~\ref{fig:polFit16}. Again,
the algebraic SDEs work satisfactorily unlike the geometric ones
for which, in particular, the sudden rise of the order parameter
appears at a larger value of $\beta$, namely $\beta_\mathrm{geo}
\simeq 6.5$. This is substantially larger than the critical
coupling, $\beta_c \simeq 5.7$. Our explanation for this behavior
is the fact that, due to the first-order nature of the transition,
there are rather sharp phase boundaries in the space of coupling
constants, $\lambda_i$. Hence, a tiny change in the couplings
(presumably well within the IMC error bars) may easily have a
large effect: by straying into the `wrong' phase the Polyakov loop
will suddenly explode or collapse. This nonlinear effect is rather
difficult to evade considering the unavoidable (if small)
instabilities of the IMC procedure. As a result, as we inevitably
increase these inaccuracies by adding more coupling we expect this
fine-tuning problem to become enhanced even further. The following
subsection will precisely address this topic.

\subsection{NN and NNN couplings}

There are (at least) two possibilities to generalise the NN
$\lambda$-action of the previous subsection. One may either extend
the NN terms to higher group representations or include
interactions of larger range, say NNN.

\begin{table}
\caption{\label{tab:consistency2}IMC consistency check for the
extended NN and NNN $\lambda$-actions including up to 11 couplings
$\lambda_i$.}
\begin{ruledtabular}
\begin{tabular}{lcccc}
 & \multicolumn{2}{c}{extended NN} & \multicolumn{2}{c}{NN + NNN} \\
  $i$ & $\lambda_{i,\mathrm{input}}$ & $\lambda_{i,\mathrm{IMC}}$ &
  $\lambda_{i,\mathrm{input}}$ & $\lambda_{i,\mathrm{IMC}}$ \\ \hline
  $1$ & $-0.0050$ & $-0.0050(1)$ & $-0.0400$ & $-0.0400(1)$ \\
  $2$ & $0.0100$ & $0.0100(1)$ & $0.0100$ & $0.0099(1)$ \\
  $3$ & $-0.0150$ & $-0.0151(2)$ & $-0.0200$ & $-0.0201(1)$ \\
  $4$ & $0.0070$ & $0.0070(1)$ & $0.0300$ & $0.0300(1)$ \\
  $5$ & $-0.0080$ & $-0.0080(5)$ & $0.0050$ & $0.0052(3)$ \\
  $6$ & $0.0090$ & $0.0091(1)$ & & \\
  $7$ & $0.0030$ & $0.0030(1)$ & & \\
  $8$ & $-0.0030$ & $-0.0030(1)$ & & \\
  $9$ & $0.0080$ & $0.0081(1)$ & & \\
  $10$ & $-0.0060$ & $-0.0058(1)$ & & \\
  $11$ & $0.0020$ & $0.0020(2)$ & & \\
  $12$ & & & $-0.0020$ & $-0.0021(1)$ \\
  $13$ & & & $0.0060$ & $0.0060(1)$ \\
  $14$ & & & $-0.0030$ & $-0.0030(1)$ \\
  $15$ & & & $0.0040$ & $0.0039(1)$ \\
  $16$ & & & $-0.0070$ & $-0.0070(1)$ \\
\end{tabular}
\end{ruledtabular}
\end{table}

The new NN terms we will add are of strong-coupling order
$\beta^{3\Nt}$, the additional NNN terms of order $\beta^{2\Nt}$.
As we are working at large $\beta$ one cannot predict which ones
are going to be more important. Rather, this will be one of the
questions to be considered in what follows.

As before we have first tested the consistency of our SDEs. Table
\ref{tab:consistency2} shows once again that even for a total of
the order of ten couplings the method works well: IMC output
reproduces input for the effective theory. The empty input entries
in Table \ref{tab:consistency2} correspond to vanishing couplings.
For a few sample couplings we have checked that vanishing input
correctly entails vanishing output as well.

\begin{figure}[h]
\includegraphics{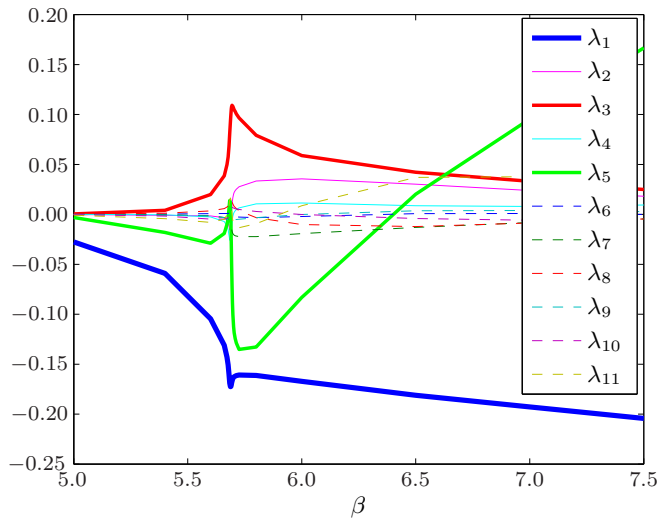}
\caption{\label{fig:couplingsNN} Behavior of the couplings
$\lambda_1, \ldots, \lambda_{11}$ (appearing in the extended NN
$\lambda$-action) as a function of the Wilson coupling $\beta$.
(IMC based on algebraic SDEs.)}
\end{figure}

\begin{figure}[h]
\includegraphics{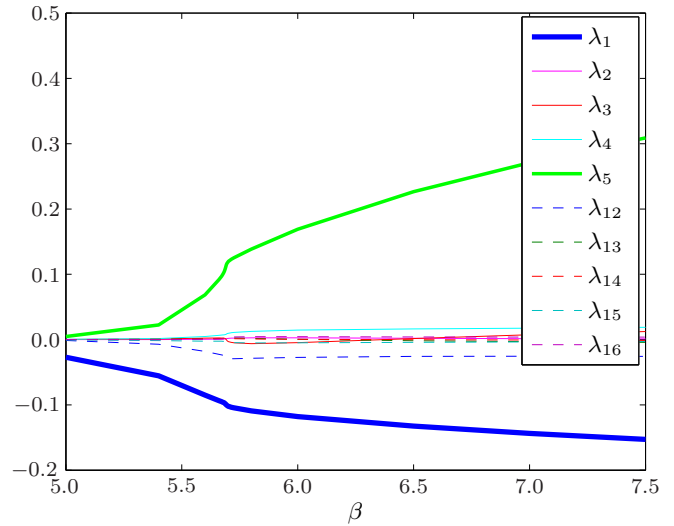}
\caption{\label{fig:couplingsP} Behavior of the couplings
$\lambda_1, \ldots, \lambda_5, \lambda_{12}, \ldots ,
\lambda_{16}$ (appearing in the NN+NNN $\lambda$-action) as a
function of the Wilson coupling $\beta$. (IMC based on algebraic
SDEs.)}
\end{figure}

To determine the behavior of the effective couplings $\lambda_i$,
$i = 1, \ldots, 16$, as a function of the Wilson coupling $\beta$
we have used a set of $4\times 10^6$ configurations per $\beta$ on
a $16^3\!\times\!4$-lattice. This amounts to $5\times10^4$
($10^3$) uncorrelated configurations far away from (close to) the
phase transition. The IMC results are shown in Figs.\
\ref{fig:couplingsNN} and \ref{fig:couplingsP} displaying the
effective couplings as functions of $\beta$. Again we note that
the fundamental and octet potential couplings ($\lambda_1$ and
$\lambda_5$) dominate in size. More important from a principal
point of view, however, is the observation that the behavior of
coupling constants is very sensitive to the choice of operators.
Let us compare, for instance, the coupling $\lambda_3$ in the two
Figs.\ \ref{fig:couplingsNN} and \ref{fig:couplingsP}. For the
extended NN ansatz (Fig.\ \ref{fig:couplingsNN}) it is comparable
in magnitude with $\lambda_1$ and $\lambda_5$ and behaves
similarly as for the simple NN ansatz of
Fig.~\ref{fig:couplings4}. However, as soon as we include NNN
operators its magnitude drops by 100\% and its behavior changes
drastically (Fig.\ \ref{fig:couplingsP}). The latter is also true
quite significantly for the octet coupling $\lambda_5$. This
clearly signals an instability of the IMC methods, at least as far
as the determination of the couplings is concerned.
%
%A further investigation not shown here leads to the fact that this
%qualitative behavior is independent of the special normalisation
%scheme while the quantitative relations are normalisation dependent.

\begin{figure}[b]
\includegraphics{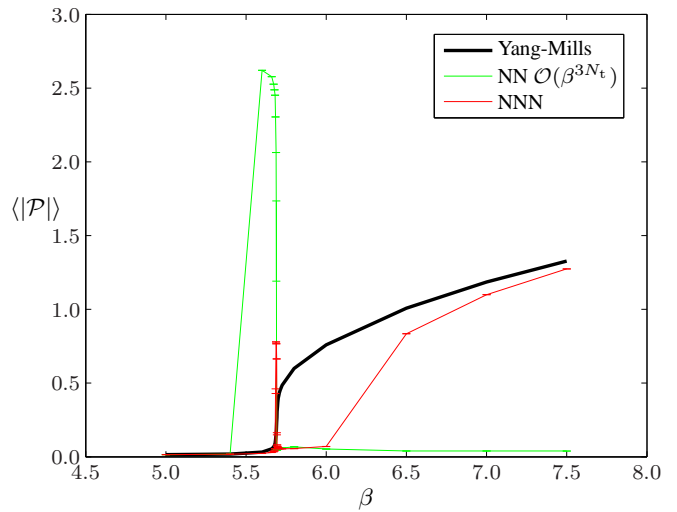}
\caption{\label{fig:polNextOrder} Comparison between Yang-Mills
and effective Polyakov loops for the extended NN-action (couplings
$\lambda_1, \ldots, \lambda_{11}$) and the NN+NNN action
(couplings $\lambda_1, \ldots, \lambda_5, \lambda_{12}, \ldots ,
\lambda_{16}$).}
\end{figure}

Nevertheless, it might still be possible that largely different
sets of couplings lead to more or less identical behavior of
observables. Comparing the behavior of the Polyakov loop in the
effective and Yang-Mills theories rules out this possibility. As
Fig.~\ref{fig:polNextOrder} shows the Polyakov loop when
calculated in the effective models is extremely sensitive to the
choice of operators and the value of $\beta$ around $\beta_c$. For
both choices of SDEs the effective order parameter significantly
overshoots the Yang-Mills one in a small $\beta$-range near
$\beta_c$ (see the spikes in Fig.~\ref{fig:polNextOrder}). This
means that in the space of effective couplings the phase boundary
to the deconfined phase have slightly (and for a short range of
$\beta$ values) been crossed albeit with drastic effect due to the
discontinuous behavior of the Polyakov loop. We conclude that the
fine tuning problem encountered in the previous subsection indeed
becomes more severe if we include more operators (and hence
increase the instabilities of the IMC procedure). For the given
number of effective couplings (of order ten) we have not been able
to get this problem under control.

% ========================================
\section{Summary and Outlook}
\label{sec:SO}
% ========================================

In this paper we have applied the IMC method to study the finite
temperature phase transition of $SU(3)$ Yang-Mills theory. A
crucial input were novel Schwinger-Dyson equations based on
algebraic and geometrical properties of the $SU(3)$ Haar measure.
The resulting equations constitute overdetermined linear systems
for the effective couplings $\lambda_i$ which were solved
numerically via least square techniques. The method works well if
the number of couplings is sufficiently small, say of the order of
five. However, already in this case one notes a fine-tuning
problem as the behavior of the Polyakov loop depends in a very
sensitive and nonlinear way on the effective couplings. This fact
can be traced back to the discontinuities associated with the
first-order character of the phase transition. Hence, the problem
becomes more pronounced in larger volumes.

If we increase the number of effective couplings and thus,
inevitably, the instabilities in their IMC determination, the
fine-tuning problem again becomes more severe. This holds to such
an extent that we could no longer gain numerical control and hence
could no longer reproduce the Yang-Mills behavior of the Polyakov
loop in the vicinity of the critical Wilson coupling, $\beta =
\beta_c$. We believe that an improvement on this situation will
require nontrivial modifications of the IMC procedure like e.g.\
smoothening of the loop in order to avoid the unphysical spikes of
Fig.~\ref{fig:polNextOrder}. In addition, it would be interesting
to check whether Creutz's microcanonical demon method
\cite{Creutz:1983ra} mentioned in the introduction yields better
results.

We conclude, nevertheless, with the positive statement that the
IMC method does work for the first-order $SU(3)$ transition as
well if we allow for only a small number of terms in the
effective Polyakov loop actions.

% ========================================
\begin{acknowledgments}
%\label{sec:ACK}
 % ========================================
TK and CW gratefully acknowledge their scholarships by the
Konrad-Adenauer-Stiftung e.V. and the Studienstiftung des
deutschen Volkes, respectively. TH thanks his colleagues of the
Plymouth Particle Theory Group for their support.
\end{acknowledgments}

\begin{appendix}
\section{Operators for the $\boldsymbol{\lambda}$-actions}\label{app:SLLIST}

\begin{figure}
\includegraphics{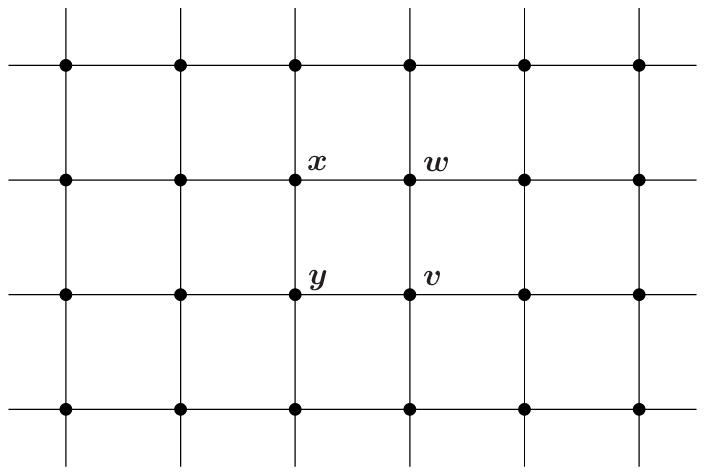}
\caption{\label{fig:latticeOnn} The neighboring relationships of
the marked sites correspond to the bracket notations
$\onN{\vcx\vcv}$, $\onN{\vcy\vcw}$, $\nN{\vcx\vcy\vcv}$,
$\nN{\vcy\vcv\vcw}$, $\nN{\vcv\vcw\vcx}$, $\nN{\vcw\vcx\vcy}$,
$\qnN{\vcx\vcy,\vcv\vcw}$ and $\qnN{\vcx\vcw,\vcy\vcv}$.}
\end{figure}

In this paper we have used up to 16 different operators appearing
in the $\lambda$-action \eqref{S-LAMBDA}:
\begin{align}
  S_1 &= \sum_{\nN{\vcx\vcy}}(\chi_{10}(\trP_\vcx)\chi_{01}(\trP_\vcy)+\coco) \; , \\
  S_2 &= \sum_{\nN{\vcx\vcy}}(\chi_{20}(\trP_\vcx)\chi_{02}(\trP_\vcy)+\coco) \; , \\
  S_3 &= \sum_{\nN{\vcx\vcy}}\chi_{11}(\trP_\vcx)\chi_{11}(\trP_\vcy) \; ,
\end{align}
\begin{align}
  S_4 &=
  \sum_{\nN{\vcx\vcy}}\begin{aligned}[t]&(\chi_{10}(\trP_\vcx)\chi_{20}(\trP_\vcy)\\
  &\qquad\phantom{\rule{0pt}{2.3ex}}+\chi_{20}(\trP_\vcx)\chi_{10}(\trP_\vcy) +\coco) \; ,\end{aligned} \\
  S_5 &= \sum_{\vcx}\chi_{11}(\trP_\vcx) \; , \\
  S_6 &= \sum_{\nN{\vcx\vcy}} \left( \chi_{30}(\trP_\vcx)
  \chi_{03}(\trP_\vcy) + \coco \right) \; , \\
  S_7 &= \sum_{\nN{\vcx\vcy}} \left( \chi_{21}(\trP_\vcx)
  \chi_{12}(\trP_\vcy) + \coco \right) \; , \\
  S_8 &= \sum_{\nN{\vcx\vcy}}\begin{aligned}[t]& ( \chi_{30}(\trP_\vcx)
  \chi_{11}(\trP_\vcy)\\
  &\qquad\phantom{\rule{0pt}{2.3ex}}+ \chi_{11}(\trP_\vcx) \chi_{30}(\trP_\vcy) + \coco ) \; ,\end{aligned} \\
  S_9 &= \sum_{\nN{\vcx\vcy}} \begin{aligned}[t]& ( \chi_{21}(\trP_\vcx)
  \chi_{20}(\trP_\vcy)\\
  &\qquad\phantom{\rule{0pt}{2.3ex}}+ \chi_{20}(\trP_\vcx) \chi_{21}(\trP_\vcy) + \coco ) \; ,\end{aligned}  \\
  S_{10} &= \sum_{\nN{\vcx\vcy}} \begin{aligned}[t]&( \chi_{21}(\trP_\vcx)
  \chi_{01}(\trP_\vcy)\\
  &\qquad\phantom{\rule{0pt}{2.3ex}}+ \chi_{01}(\trP_\vcx) \chi_{21}(\trP_\vcy)
  + \coco ) \; , \end{aligned} \\
  S_{11} &= \sum_{\vcx} \left(\chi_{30}(\trP_\vcx) + \coco\right) \; , \\
  S_{12} &= \sum_{\onN{\vcx\vcz}}(\chi_{10}(\trP_\vcx)\chi_{01}(\trP_\vcz) +\coco ) \; , \\
  S_{13} &=
  \sum_{\nN{\vcx\vcy\vcz}}(\chi_{10}(\trP_\vcx)\chi_{01}(\trP_\vcz)
  +\coco )\chi_{11}(\trP_\vcy) \; , \\
  S_{14} &=
  \sum_{\nN{\vcx\vcy\vcz}}(\chi_{10}(\trP_\vcx)\chi_{02}(\trP_\vcy)\chi_{10}(\trP_\vcz)
  +\coco ) \; , \\
  S_{15} &=
  \sum_{\nN{\vcx\vcy\vcz}}(\chi_{10}(\trP_\vcx)\chi_{10}(\trP_\vcy)\chi_{10}(\trP_\vcz)
  +\coco ) \; , \\
  S_{16} &= 
  \sum_{\qnN{\vcx\vcy,\vcv\vcw}}\begin{aligned}[t] &(\chi_{10}(\trP_\vcx) \chi_{01} (\trP_\vcy)+\coco)\\
  &\qquad\phantom{\rule{0pt}{2.3ex}} \cdot(\chi_{10}(\trP_\vcv)\chi_{01}
  (\trP_\vcw)+\coco) \; .\end{aligned}
\end{align}
The NN and NNN relationships are denoted in terms of brackets the
meaning of which is explained in Fig.~\ref{fig:latticeOnn}. Hence,
the operators $S_1$ to $S_{11}$ obviously describe (extended) NN
interactions, while $S_{12}$ to $S_{16}$ are NNN terms. In a
strong coupling (small-$\beta$) expansion the terms $S_1, \ldots ,
S_5$ and $S_{12}, \ldots , S_{16}$ would be of $\ord(\beta^{2\Nt})$,
the terms $S_6, \ldots , S_{11}$ of $\ord(\beta^{3\Nt})$
\cite{Wozar:2006dp}.

\section{Operators for the $\boldsymbol{\kappa}$-actions}\label{app:SKLIST}

If we extend the strong coupling NN contributions to $\ord(\beta^{3\Nt})$ the
effective action becomes a series of nine terms,
\begin{equation} \label{XNN}
  \Seff = \sum_{a=1}^9 \kappa_a I_a \; ,
\end{equation}
which is referred to as the extended NN action (as is its
$\lambda$-equivalent, see Appendix~\ref{app:LAMBDA-KAPPA} below).

If we allow for NNN interactions (which, however, do not extend
beyond single plaquettes) up to order $\ord(\beta^{2\Nt})$ we end
up with the effective action
\begin{equation} \label{NN+NNN}
  \Seff = \sum_{a\in\{1,2,3,4,10,11\}} \kappa_a I_a \; .
\end{equation}
This (and its $\lambda$-equivalent, see
Appendix~\ref{app:LAMBDA-KAPPA} below) is referred to as the
NN+NNN action.

The resulting operators are given by
\begin{align}
  I_1 &= \sum_{\nN{\vcx\vcy}}(\chi_{10}(\trP_\vcx)\chi_{01}(\trP_\vcy)+\coco) \; , \\
  I_2 &= \sum_{\nN{\vcx\vcy}}(\chi_{20}(\trP_\vcx)\chi_{02}(\trP_\vcy)+\coco) \; , \\
  I_3 &= \sum_{\nN{\vcx\vcy}}\chi_{11}(\trP_\vcx)\chi_{11}(\trP_\vcy) \; ,\\
  I_4 &= \sum_{\nN{\vcx\vcy}}(\chi_{10}(\trP_\vcx)\chi_{01}(\trP_\vcy)+\coco)^2 \; , \\
  I_5 &= \sum_{\nN{\vcx\vcy}}(\chi_{30}(\trP_\vcx)\chi_{03}(\trP_\vcy)+\coco) \; , \\
  I_6 &= \sum_{\nN{\vcx\vcy}}(\chi_{21}(\trP_\vcx)\chi_{12}(\trP_\vcy)+\coco) \; , \\
  I_7 &= 
  \sum_{\nN{\vcx\vcy}}\begin{aligned}[t]&(\chi_{10}(\trP_\vcx)\chi_{01}(\trP_\vcy)+\coco) \\
  &\qquad\phantom{\rule{0pt}{2.3ex}} \cdot(\chi_{20}(\trP_\vcx)\chi_{02}(\trP_\vcy)+\coco) \;
  ,\end{aligned} \\
  I_8 &=
  \sum_{\nN{\vcx\vcy}}\begin{aligned}[t]&(\chi_{10}(\trP_\vcx)\chi_{01}
  (\trP_\vcy)+\coco)\\ &\qquad\quad\phantom{\rule{0pt}{2.3ex}}\cdot\chi_{11}
  (\trP_\vcx)\chi_{11}(\trP_\vcy)   \; ,\end{aligned} \\
  I_9 &= \sum_{\nN{\vcx\vcy}}(\chi_{10}(\trP_\vcx)\chi_{01}(\trP_\vcy)+\coco)^3 \; , 
\end{align}
\begin{align}
  I_{10} &= 
  \sum_{\nN{\vcx\vcy\vcz}}\begin{aligned}[t]&(\chi_{10} (\trP_\vcx) \chi_{01} (\trP_\vcy)+\coco) \\
  &\qquad\phantom{\rule{0pt}{2.3ex}} \cdot(\chi_{10}
  (\trP_\vcy)\chi_{01} (\trP_\vcz)+\coco) \; , \end{aligned} \\
  I_{11} &= 
  \sum_{\qnN{\vcx\vcy,\vcv\vcw}}\begin{aligned}[t] &(\chi_{10} (\trP_\vcx)\chi_{01}
  (\trP_\vcy)+\coco) \\ &\qquad\phantom{\rule{0pt}{2.3ex}} \cdot (\chi_{10}
  (\trP_\vcv) \chi_{01}(\trP_\vcw)+\coco) \; .\end{aligned}
\end{align}
\\

\section{Linear coupling relations}\label{app:LAMBDA-KAPPA}

For the extended NN action \eqref{XNN} the relation between the
$\lambda_i$ and the $\kappa_a$ is

\begin{equation}\label{eq:couplRelation}
\begin{pmatrix}
\lambda_1 \\ \lambda_2 \\ \lambda_3 \\ \lambda_4 \\ \lambda_5 \\
\lambda_6 \\ \lambda_7 \\ \lambda_8 \\ \lambda_9 \\ \lambda_{10} \\ \lambda_{11}
\end{pmatrix} =
\begin{pmatrix}
1 & 0 & 0 & 1 & 0 & 0 & 1 & 1 & 12 \\
0 & 1 & 0 & 1 & 0 & 0 & 0 & 1 & 3 \\
0 & 0 & 1 & 2 & 0 & 0 & 2 & 0 & 8 \\
0 & 0 & 0 & 1 & 0 & 0 & 0 & 1 & 6 \\
0 & 0 & 0 & 12 & 0 & 0 & 0 & 0 & 24 \\
0 & 0 & 0 & 0 & 1 & 0 & 1 & 0 & 1 \\
0 & 0 & 0 & 0 & 0 & 1 & 1 & 1 & 3 \\
0 & 0 & 0 & 0 & 0 & 0 & 1 & 0 & 2 \\
0 & 0 & 0 & 0 & 0 & 0 & 0 & 1 & 3 \\
0 & 0 & 0 & 0 & 0 & 0 & 1 & 1 & 6 \\
0 & 0 & 0 & 0 & 0 & 0 & 0 & 0 & 6
\end{pmatrix}
\begin{pmatrix}
\kappa_1 \\ \kappa_2 \\ \kappa_3 \\ \kappa_4 \\ \kappa_5 \\ \kappa_6 \\ \kappa_7 \\ \kappa_8 \\
\kappa_9 \end{pmatrix} \; .
\end{equation}

The analogous relation for the NN+NNN action \eqref{NN+NNN} is

\begin{equation}
\begin{pmatrix}
\lambda_1 \\ \lambda_2 \\ \lambda_3 \\ \lambda_4 \\ \lambda_5 \\
\lambda_{12} \\ \lambda_{13} \\ \lambda_{14} \\ \lambda_{15} \\ \lambda_{16}
\end{pmatrix} =
\begin{pmatrix}
1 & 0 & 0 & 1 & 0 & 0 \\
0 & 1 & 0 & 1 & 0 & 0 \\
0 & 0 & 1 & 2 & 0 & 0 \\
0 & 0 & 0 & 1 & 0 & 0 \\
0 & 0 & 0 & 12 & 0 & 0 \\
0 & 0 & 0 & 0 & 2 & 0 \\
0 & 0 & 0 & 0 & 1 & 0 \\
0 & 0 & 0 & 0 & 1 & 0 \\
0 & 0 & 0 & 0 & 1 & 0 \\
0 & 0 & 0 & 0 & 0 & 1
\end{pmatrix}
\begin{pmatrix}
\kappa_1 \\ \kappa_2 \\ \kappa_3 \\ \kappa_4 \\ \kappa_{10} \\ \kappa_{11}
\end{pmatrix} \; .
\end{equation}

\end{appendix}

%and here comes the bbl-file
%======================================

\end{document}